\documentclass[11pt,english,amsmath]{revtex4}
\usepackage[T1]{fontenc}
\usepackage[latin9]{inputenc}
\usepackage{amsmath}
\usepackage{esint}

\makeatletter
\@ifundefined{textcolor}{}
{%
 \definecolor{BLACK}{gray}{0}
 \definecolor{WHITE}{gray}{1}
 \definecolor{RED}{rgb}{1,0,0}
 \definecolor{GREEN}{rgb}{0,1,0}
 \definecolor{BLUE}{rgb}{0,0,1}
 \definecolor{CYAN}{cmyk}{1,0,0,0}
 \definecolor{MAGENTA}{cmyk}{0,1,0,0}
 \definecolor{YELLOW}{cmyk}{0,0,1,0}
 }

\makeatother

\usepackage{babel}

\begin{document}

\title{Green's Matrix for a Second Order Self-Adjoint Matrix Differential
Operator}

\author{Tahsin Ça\u{g}r\i{} \c{S}i\c{s}man}

\email{sisman@metu.edu.tr}

\affiliation{Department of Physics,\\
 Middle East Technical University, 06531, Ankara, Turkey}

\author{Bayram Tekin}

\email{btekin@metu.edu.tr}

\affiliation{Department of Physics,\\
 Middle East Technical University, 06531, Ankara, Turkey}

\date{\today}
\begin{abstract}
A systematic construction of the Green's matrix for a second order,
self-adjoint matrix differential operator from the linearly independent
solutions of the corresponding homogeneous differential equation set
is carried out. We follow the general approach of extracting the Green's
matrix from the Green's matrix of the corresponding first order system.
This construction is required in the cases where the differential
equation set cannot be turned to an algebraic equation set via transform
techniques.
\end{abstract}
\maketitle

\section{Introduction}

In physics, matrix differential operators acting on vector functions
appear in many different contexts from classical electromagnetism
to quantum field theory. Green's matrices of these operators are needed
because of their own physical interpretation as propagators in quantum
field theory, or in order to find the solutions of the corresponding
non-homogeneous differential equation set.

In most of the cases, Green's matrices are obtained by using the Fourier
transform technique or by using eigenfunction expansions which turn
the differential equation set to an algebraic one. However, these
techniques are not applicable in some circumstances such as for the
matrix differential operator appearing in the 1+1 dimensional Abelian-Higgs
model \cite{Baacke and Daiber,Baacke 2008}. In this model, when small
field fluctuations around classical field configurations are investigated,
Lagrangian of the theory, which is second order in fluctuations, involve
a $4\times4$ matrix differential operator. Diagonal entries of this
operator are differential operators of the modified Bessel type. Off-diagonal
{}``potential'' terms are functions of classical field configurations
which are only available as discrete numeric data for generic values
of parameters in the theory. Green's matrix of this operator is required
in calculating the functional determinant of the operator which gives
one-loop corrections about a classical solution such as an instanton
(For a nice account of functional determinants using Gel'fand-Yaglom
technique see \cite{Dunne 2008}). Green's function technique is one
of the methods used in such a determinant calculation. Due to the
existence of the modified Bessel type operators and the discrete numerical
data, it is not possible to apply Fourier transform and eigenfunction
expansion in this case. However, in a numerical study, it is relatively
easy to obtain linearly independent solutions of the corresponding
homogeneous differential equation. Therefore, construction of the
Green's matrix from these solutions as in the case of the single differential
operator is required.

It is worth considering the underlying physical problem in some detail
in order to see why one encounters a matrix differential operator
which is hard to handle. In \cite{tHooft}, 't Hooft studied the one-loop
tunneling amplitude in the background of a Yang-Mills instanton for
a theory which contains \emph{massless} scalar and fermion fields
. In this calculation, field fluctuations do not couple, and the functional
determinants of \emph{single} differential operators are calculated.
Dunne \emph{et} \emph{al.}$\!$ \cite{Dunne 2005a,Dunne 2005b} extended
the instanton determinant calculation to the arbitrary quark mass
case. 't Hooft \cite{tHooft} pointed out that in order to remove
the infrared divergence of the theory, one needs to introduce the
Higgs field. However, due to simple scaling arguments, there is no
instanton solution in this case. It is still viable to do the calculations
in the same instanton background, but with certain instanton size,
since it was shown in \cite{tHooft} and in \cite{Affleck}, with
a more elaborate discussion, that the Higgs particle can be taken
as approximately massless. However, if one studies the effect of the
quantum fluctuations around an instanton background with non-trivial
Higgs field configuration, as in the case of the $1+1$ Abelian-Higgs
model, field fluctuations \emph{do} couple to each other and one needs
to struggle with the functional determinant calculation of a matrix
differential operator.

In the physics literature, a\emph{ general construction} of the Green's
matrix from the set of solutions of the corresponding homogeneous
differential equations does not seem to exist. Among the standard
references of mathematical physics, only Courant and Hilbert \cite{Courant and Hilbert}
involves a short discussion on the properties of the Green's matrix
({}``tensor'' as called in \cite{Courant and Hilbert}) without
a construction. Baacke \cite{Baacke 1997} gave a construction for
a specific matrix differential operator in a heuristic way. He studied
one-loop effects in various field theories using Green's matrices
found by this construction \cite{Baacke and Daiber,Baacke 2008,Baacke 1997,Baacke 1990}.
In \cite{Zuhuravlev et al}, again there is a construction for a specific
operator with the main emphasis on the boundary conditions of the
underlying physical system which is a magnetic multilayer structure.

In the mathematics literature, construction of Green's matrix for
a second order, self-adjoint differential operator from the solutions
of the corresponding homogeneous differential equation set does exist.
Naimark \cite{Naimark} studied Green's matrices of the $n^{\text{th}}$
order, linear matrix differential operators for general homogeneous
boundary conditions relating vector function and its derivatives up
to the $\left(n-1\right)^{\text{th}}$ order at the boundaries. He
gave the Green's matrix form for this general system and outlines
a way to prove his result. These results and analysis are too general,
so a special study of the physically relevant case, which is the second
order and self-adjoint operator, is still valuable. Let us mention
several other related works. Bhagat \cite{Bhagat a,Bhagat b} worked
on the case of second order, self-adjoint, $2\times2$ matrix differential
operator. Heimes \cite{Heimes} gave an analog of our result for second
order, linear systems without any explicit construction, and only
mentioned that the proper method is to transform the second order
system to first order system. Jodar \cite{Jodar} worked on an algebraic
construction which may not work on every case.

In this paper, we consider a generic second order, self-adjoint %
\footnote{Self-adjointness requires symmetry of the matrix operator besides
the self-adjointness of the differential operators on the diagonals.\cite{Naimark}%
} $n\times n$ matrix differential operator of the form\begin{equation}
\mathbf{M}_{x}=\left[\begin{array}{cccc}
M_{11,x} & V_{12}\left(x\right) & \dots & V_{1n}\left(x\right)\\
V_{12}\left(x\right) & M_{22,x} & \dots & V_{2n}\left(x\right)\\
\vdots & \vdots & \ddots & \vdots\\
V_{1n}\left(x\right) & V_{2n}\left(x\right) & \dots & M_{nn,x}\end{array}\right],\label{eq:mtrx_diff_op}\end{equation}
where the diagonal entries are of the form\[
M_{ii,x}\equiv\left[\frac{d}{dx}\left(p_{i}\left(x\right)\frac{d}{dx}\right)+q_{i}\left(x\right)\right],\]
and the off-diagonal entries $V_{ij}=V_{ji}$ are just continuous
functions. First of all, we develop the properties of the Green's
matrix in Section \ref{sec:Properties-of-Green's}. Section \ref{sec:Construction-of-Green's}
is devoted to the construction of the Green's matrix. Construction
is carried out in two ways. In the first way, Green's matrix of second
order system is extracted from the Green's matrix of the corresponding
first order system. In this construction,\emph{ }general approach
developed in Cole \cite{Cole} is followed %
\footnote{Also, Reid \cite{Reid} presents the same analysis which is followed
by Heimes\cite{Heimes}.%
}. This construction relies on the same basic idea as the constructions
of \cite{Zuhuravlev et al} and \cite{Heimes}. In the second way
of construction, we start with a guess on the form of the Green's
matrix, which is along the line of \cite{Baacke 1997}\emph{.}

To fix the notation, let us note that small bold letters, \emph{e.g.$\!$}
$\mathbf{y}$, represent vectors; capital bold letters, \emph{e.g.$\!$}
$\mathbf{G}$ represent matrices. $x$ appearing as an index refers
to a differential operator such as $\mathbf{M}_{x}$. Repeated indices
on different matrices are to be summed over, unless otherwise stated.
Repeated indices on a single matrix refer to a diagonal element. And
a word about the nomenclature: to distinguish the Green's \emph{function}
of a single differential operator, we choose the name Green's \emph{matrix}
for coupled equations.

\section{Properties of the Green's Matrix\label{sec:Properties-of-Green's}}

Let's consider a linear, second order, coupled differential equation
set;\begin{equation}
\mathbf{M}_{x}\mathbf{y}\left(x\right)=\mathbf{h}\left(x\right),\qquad x\in\left[a,b\right],\label{eq:scnd_ord_sys}\end{equation}
where $\mathbf{y}$ , $\mathbf{h}$ are $n$ dimensional vector functions,
and $\mathbf{M}_{x}$ is $n\times n$ dimensional, self-adjoint matrix
differential operator of the form (\ref{eq:mtrx_diff_op}). Green's
matrix, $\mathbf{G}\left(x,t\right)$ of the differential operator,
$\mathbf{M}_{x}$ can be defined with the formal solution\begin{equation}
\mathbf{y}\left(x\right)=\int_{a}^{b}dt\,\mathbf{G}\left(x,t\right)\mathbf{h}\left(t\right),\label{eq:formal_soln_second}\end{equation}
where $\mathbf{G}\left(x,t\right)$ is $n\times n$ matrix. 

In this paper, homogeneous boundary conditions are considered:\[
\mathbf{y}\left(a\right)=\mathbf{o},\quad\mathbf{y}\left(b\right)=\mathbf{o},\]
where $\mathbf{o}$ is the $n$ dimensional zero vector. These boundary
conditions impose the following conditions on the Green's matrix;\[
\mathbf{G}\left(a,t\right)=\mathbf{O},\quad\mathbf{G}\left(b,t\right)=\mathbf{O},\]
where $\mathbf{O}$ is $n\times n$ zero matrix.

The formal solution of the differential equation set implies\begin{equation}
M_{x,ij}G_{jk}\left(x,t\right)=\delta_{ik}\delta\left(x-t\right),\label{eq:defn_delta_mtrx}\end{equation}
where $i$, $j$, $k$ indices run from $1$ to $n$. This relation
is an equality of distributions. Integrals of these distributions
with a test function yield further properties of the Green's matrix.
In obtaining these properties, $i=k$ and $i\neq k$ cases will be
investigated separately.

\textbf{\underbar{For $i\neq k$:}}

$k^{\text{th}}$ column of the Green's matrix is a solution of the
homogeneous differential equations except the $k^{\text{th}}$ equation,
as implied by\[
M_{x,ij}G_{jk}\left(x,t\right)=0.\]
Each differential equation involves a term of the form\[
M_{x,ii}G_{ik}\left(x,t\right)=\left[\frac{d}{dx}\left(p_{i}\left(x\right)\frac{d}{dx}\right)+q_{i}\left(x\right)\right]G_{ik}\left(x,t\right)\]
where there is no summation on $i$. In order to satisfy these $\left(n-1\right)$
homogeneous differential equations, first and second order derivatives
in the above term should not yield any singularities, since the other
terms of the differential equation contain continuous potentials.
Thus, elements of the $k^{\text{th}}$ column should be continuous
and should have continuous first derivatives for any $x\in\left[a,b\right]$,
except $G_{kk}\left(x,t\right)$ which is investigated in $i=k$ case
below.

\textbf{\underbar{For $i=k$:}}

$k^{\text{th}}$ column of the Green's matrix is a solution of the
$k^{\text{th}}$ homogeneous differential equation for $x\in\left[a,t\right)\cup\left(t,b\right]$,
as implied by\[
M_{x,kj}G_{jk}\left(x,t\right)=\delta\left(x-t\right),\]
where there is no summation on $k$. This equation contains the term\[
M_{x,kk}G_{kk}\left(x,t\right)=\left[\frac{d}{dx}\left(p_{k}\left(x\right)\frac{d}{dx}\right)+q_{k}\left(x\right)\right]G_{kk}\left(x,t\right).\]
Since\[
M_{x,kj}G_{jk}\left(x,t\right)=0,\]
for $x\ne t$, $G_{kk}\left(x,t\right)$ should be continuous and
has continuous first derivatives for points other than $x=t$.

Let's consider the behavior at $x=t$. Since all elements of the $k^{\text{th}}$
column other than $G_{kk}\left(x,t\right)$ are continuous at $x=t$,
Dirac delta behavior comes from $G_{kk}\left(x,t\right)$. A discontinuity
in $G_{kk}\left(x,t\right)$ yields a more severe singularity than
Dirac delta upon taking the second derivative. Thus, $G_{kk}\left(x,t\right)$
is continuous at $x=t$ and the first derivative of $G_{kk}\left(x,t\right)$
has the usual discontinuity\[
\underset{\epsilon\rightarrow0}{\lim}\left.\frac{d}{dx}G_{ii}\left(x,t\right)\right|_{t-\epsilon}^{t+\epsilon}=\frac{1}{p_{i}\left(t\right)}.\]

As in the case of the Green's function for a single differential operator,
self-adjointness of $\mathbf{M}_{x}$ and the homogeneous boundary
conditions yield a symmetry property for the Green's matrix:\begin{multline*}
\left(\mathbf{G}^{T}\right)_{li}\left(x,x_{2}\right)M_{x,ij}G_{jk}\left(x,x_{1}\right)=\left(\mathbf{G}^{T}\right)_{li}\left(x,x_{2}\right)\delta_{ik}\delta\left(x-x_{1}\right),\\
\Rightarrow G_{il}\left(x,x_{2}\right)M_{x,ij}G_{jk}\left(x,x_{1}\right)=G_{kl}\left(x,x_{2}\right)\delta\left(x-x_{1}\right),\end{multline*}
and\begin{multline*}
\left(\mathbf{G}^{T}\right)_{ki}\left(x,x_{1}\right)M_{x,ij}G_{jl}\left(x,x_{2}\right)=\left(\mathbf{G}^{T}\right)_{ki}\left(x,x_{1}\right)\delta_{il}\delta\left(x-x_{2}\right),\\
\Rightarrow G_{ik}\left(x,x_{1}\right)M_{x,ij}G_{jl}\left(x,x_{2}\right)=G_{lk}\left(x,x_{1}\right)\delta\left(x-x_{2}\right).\end{multline*}
After integrating the above two equations over the interval $\left[a,b\right]$
and subtracting them side by side, one obtains \[
\int_{a}^{b}\left[G_{il}\left(x,x_{2}\right)M_{x,ij}G_{jk}\left(x,x_{1}\right)-G_{ik}\left(x,x_{1}\right)M_{x,ij}G_{jl}\left(x,x_{2}\right)\right]\, dx=G_{kl}\left(x_{1},x_{2}\right)-G_{lk}\left(x_{2},x_{1}\right).\]
Calculating the left-hand side:

\textbf{\underbar{For $i=j$:}}\begin{multline*}
\underset{i}{\sum}\int_{a}^{b}\left[G_{il}\left(x,x_{2}\right)\frac{d}{dx}\left(p_{i}\left(x\right)\frac{d}{dx}G_{ik}\left(x,x_{1}\right)\right)-G_{ik}\left(x,x_{1}\right)\frac{d}{dx}\left(p_{i}\left(x\right)\frac{d}{dx}G_{il}\left(x,x_{2}\right)\right)\right]\, dx.\end{multline*}
Adding and subtracting $p_{i}\left(x\right)\frac{d}{dx}G_{il}\left(x,x_{2}\right)\frac{d}{dx}G_{ik}\left(x,x_{1}\right)$
yield \begin{multline*}
\underset{i}{\sum}\int_{a}^{b}\left[\frac{d}{dx}\left(G_{il}\left(x,x_{2}\right)p_{i}\left(x\right)\frac{d}{dx}G_{ik}\left(x,x_{1}\right)\right)-\frac{d}{dx}\left(G_{ik}\left(x,x_{1}\right)p_{i}\left(x\right)\frac{d}{dx}G_{il}\left(x,x_{2}\right)\right)\right]\, dx.\end{multline*}
After the integration, one obtains\[
\underset{i}{\sum}\left[\left(G_{il}\left(x,x_{2}\right)p_{i}\left(x\right)\frac{d}{dx}G_{ik}\left(x,x_{1}\right)\right)-\left(G_{ik}\left(x,x_{1}\right)p_{i}\left(x\right)\frac{d}{dx}G_{il}\left(x,x_{2}\right)\right)\right]_{x=a}^{x=b}=0,\]
from $G_{jk}\left(a,x^{\prime}\right)=0,\, G_{jk}\left(b,x^{\prime}\right)=0$
for any $j$ and $k$.

\textbf{\underbar{For $i\neq j$:}}\[
\underset{i,j;i\neq j}{\sum}\int_{a}^{b}\left[G_{il}\left(x,x_{2}\right)M_{x,ij}G_{jk}\left(x,x_{1}\right)-G_{ik}\left(x,x_{1}\right)M_{x,ij}G_{jl}\left(x,x_{2}\right)\right]\, dx,\]
contains terms like\begin{multline*}
\left\{ \int_{a}^{b}\left[G_{1l}\left(x,x_{2}\right)M_{x,12}G_{2k}\left(x,x_{1}\right)-G_{1k}\left(x,x_{1}\right)M_{x,12}G_{2l}\left(x,x_{2}\right)\right]\, dx\right\} +\\
\left\{ \int_{a}^{b}\left[G_{2l}\left(x,x_{2}\right)M_{x,21}G_{1k}\left(x,x_{1}\right)-G_{2k}\left(x,x_{1}\right)M_{x,21}G_{1l}\left(x,x_{2}\right)\right]\, dx\right\} ,\end{multline*}
which vanish, since the matrix differential operator, $\mathbf{M}_{x}$
is symmetric.

Thus, one obtains the symmetry property of the Green's matrix:\[
G_{kl}\left(x_{1},x_{2}\right)=G_{lk}\left(x_{2},x_{1}\right).\]

As a result, Green's matrix of a second order self-adjoint matrix
differential operator satisfies (\ref{eq:defn_delta_mtrx}) and the
homogeneous boundary conditions. Rewriting them together, we have:\[
M_{x,ij}G_{jk}\left(x,x^{\prime}\right)=\delta_{ik}\delta\left(x-x^{\prime}\right);\quad G_{jk}\left(a,x^{\prime}\right)=0,\, G_{jk}\left(b,x^{\prime}\right)=0.\]
Properties of the Green's matrix developed in this section can be
summarized as:
\begin{itemize}
\item $k^{\text{th}}$ column of the Green's matrix satisfies the homogeneous
differential equations except at one point $x=t$ for equation $i=k$.\[
M_{x,ij}G_{jk}\left(x,t\right)=0,\quad x\in\begin{cases}
\left[a,t\right)\cup\left(t,b\right], & i=k,\\
\left[a,b\right], & i\neq k.\end{cases}\]

\item Green's matrix is continuous at $x=t$.
\item Derivative of the Green's matrix at point $x=t$ is continuous for
the off-diagonal elements and has a jump of $1/p_{i}\left(t\right)$
for diagonal elements.\[
\underset{\epsilon\rightarrow0}{\lim}\left.\frac{d}{dx}G_{ij}\left(x,t\right)\right|_{t-\epsilon}^{t+\epsilon}=\begin{cases}
\frac{1}{p_{i}\left(t\right)}, & i=j,\\
0 & i\ne j\end{cases}\]

\item Green's matrix has the the following symmetry:\[
\mathbf{G}\left(x,t\right)=\mathbf{G}^{T}\left(t,x\right).\]

\end{itemize}
These properties are also given in \cite{Courant and Hilbert}.

\section{Construction of Green's Matrix\label{sec:Construction-of-Green's}}

A standard way of constructing the Green's function for a second order,
linear, self-adjoint differential operator\[
L_{x}\equiv\frac{d}{dx}\left(p\left(x\right)\frac{d}{dx}\right)+q\left(x\right),\]
defined in $\left[a,b\right]$ is to use the two linearly independent
solutions of homogeneous differential equation satisfying,\begin{align*}
L_{x}u\left(x\right) & =0,\qquad u\left(a\right)=0,\\
L_{x}v\left(x\right) & =0,\qquad v\left(b\right)=0.\end{align*}
The motivation for such a construction follows the observation of
two points. First, Green's function satisfies the homogeneous differential
equation, except at $x=t$. Second point is the correspondence between
the derivative property of Green's function;\[
\underset{\epsilon\rightarrow0}{\lim}\left.\frac{d}{dx}G\left(x,t\right)\right|_{t-\epsilon}^{t+\epsilon}=\frac{1}{p\left(t\right)},\]
and the Wronskian of the solutions $u$ and $v$;\begin{equation}
W\left(u,v\right)=uv^{\prime}-vu^{\prime}=\frac{\mbox{constant}}{p}.\label{eq:wronskian}\end{equation}

Since the columns of our Green's matrix satisfy the homogeneous differential
equation set except at one point $x=t$, it is suggestive that the
Green's matrix can be constructed from the solutions of the homogeneous
differential equation. In this section, this construction will be
given. In Section \ref{sub:Analogs-of-Wronskian}, analogs of the
Wronskian, (\ref{eq:wronskian}), are obtained. A direct approach
for constructing the Green's matrix involves first transforming the
second order differential equation set to a first order differential
equation set. Then, Green's matrix of the second order set is extracted
from the Green's matrix of the first order set. This approach is handled
in Section \ref{sub:Green-mtrx-2-from-1}.

\subsection{Analogs of the Wronskian\label{sub:Analogs-of-Wronskian}}

In general, $2n$ linearly independent solutions of\[
\mathbf{M}_{x}\mathbf{y}\left(x\right)=\mathbf{o},\]
can be (re)defined in such a way that $n$ of them satisfy the left
boundary condition, and the others satisfy the right boundary condition.
Let's call them as $\mathbf{u}^{\alpha}$ and $\mathbf{v}^{\beta}$,
respectively, satisfying\[
\mathbf{u}^{\alpha}\left(a\right)=\mathbf{o},\quad\mathbf{v}^{\beta}\left(b\right)=\mathbf{o},\]
where Greek superscripts label the solutions. 

Following the similar steps leading to (\ref{eq:wronskian}), it is
possible to obtain analog relations for the matrix differential operator
case. One can write\begin{align*}
u_{i}^{\alpha}\left(x\right)M_{x,ij}v_{j}^{\beta}\left(x\right) & =0,\\
v_{i}^{\beta}\left(x\right)M_{x,ij}u_{j}^{\alpha}\left(x\right) & =0.\end{align*}
Subtracting side by side yields ($\alpha$ and $\beta$ superscripts
are suppressed since the equation holds for every $\alpha$ and $\beta$.
Also, $x$ dependence of the solutions are not explicitly shown up
until the final result.) \[
u_{i}M_{x,ij}v_{j}-v_{i}M_{x,ij}u_{j}=0.\]
Since the matrix differential operator is symmetric, terms like $u_{1}M_{12}v_{2}$
and $v_{2}M_{21}u_{1}$ cancel each other. After these cancellations,
one obtains\begin{align*}
\underset{i}{\sum}\left(u_{i}M_{x,ii}v_{i}-v_{i}M_{x,ii}u_{i}\right) & =0\Rightarrow\underset{i}{\sum}\left[u_{i}\left(p_{i}v_{i}^{\prime}\right)^{\prime}-v_{i}\left(p_{i}u_{i}^{\prime}\right)^{\prime}\right]=0,\\
 & \phantom{=0}\Rightarrow\underset{i}{\sum}\left[\left(u_{i}p_{i}v_{i}^{\prime}\right)^{\prime}-\left(v_{i}p_{i}u_{i}^{\prime}\right)^{\prime}\right]=0,\\
 & \phantom{=0}\Rightarrow\underset{i}{\sum}\left(u_{i}p_{i}v_{i}^{\prime}-v_{i}p_{i}u_{i}^{\prime}\right)=\mbox{constant}.\end{align*}

After putting the superscripts which label the solutions, and showing
the explicit $x$ dependence of solutions, one ends up with \begin{equation}
\underset{i}{\sum}p_{i}\left(x\right)\left(u_{i}^{\alpha}\left(x\right)\frac{d}{dx}v_{i}^{\beta}\left(x\right)-v_{i}^{\beta}\left(x\right)\frac{d}{dx}u_{i}^{\alpha}\left(x\right)\right)=C^{\alpha\beta},\label{eq:1/p_rltn}\end{equation}
where $C^{\alpha\beta}$ are constants. Matrix form of this equation
is\begin{equation}
\mathbf{U}^{T}\left(x\right)\mathbf{P}\left(x\right)\mathbf{V}^{\prime}\left(x\right)-\left(\mathbf{U^{\prime}}\right)^{T}\left(x\right)\mathbf{P}\left(x\right)\mathbf{V}\left(x\right)=\mathbf{C},\label{eq:W_analog1}\end{equation}
where $\mathbf{U}$ and $\mathbf{V}$ are $n\times n$ matrices whose
columns are $\mathbf{u}^{\alpha}$ and $\mathbf{v}^{\beta}$ vectors,
respectively; and $\mathbf{P}$ matrix is defined as\[
\mathbf{P}\left(x\right)\equiv\text{diag}\left[p_{1}\left(x\right),p_{2}\left(x\right),\dots,p_{n}\left(x\right)\right].\]

Other two relations that can be derived similarly are,\begin{align}
\mathbf{U}^{T}\left(x\right)\mathbf{P}\left(x\right)\mathbf{U}^{\prime}\left(x\right)-\left(\mathbf{U^{\prime}}\right)^{T}\left(x\right)\mathbf{P}\left(x\right)\mathbf{U}\left(x\right) & =\mathbf{O},\label{eq:W_analog2}\\
\mathbf{V}^{T}\left(x\right)\mathbf{P}\left(x\right)\mathbf{V}^{\prime}\left(x\right)-\left(\mathbf{V^{\prime}}\right)^{T}\left(x\right)\mathbf{P}\left(x\right)\mathbf{V}\left(x\right) & =\mathbf{O}.\label{eq:W_analog3}\end{align}
Here, using the boundary conditions $\mathbf{u}^{\alpha}\left(a\right)=\mathbf{o}$
and $\mathbf{v}^{\beta}\left(b\right)=\mathbf{o}$ in (\ref{eq:W_analog2})
and (\ref{eq:W_analog3}), respectively, one finds that the constant
matrices on the right-hand sides equal to zero. Rearranging these
equations yields symmetric matrices\begin{align}
\mathbf{P}\mathbf{U}^{\prime}\mathbf{U}^{-1} & =\left(\mathbf{U^{\prime}}\mathbf{U}^{-1}\mathbf{P}\right)^{T},\label{eq:sym_form1}\\
\mathbf{P}\mathbf{V^{\prime}}\mathbf{V}^{-1} & =\left(\mathbf{V^{\prime}}\mathbf{V}^{-1}\mathbf{P}\right)^{T},\label{eq:sym_form2}\end{align}
in $\left(a,b\right)$, since the $\mathbf{P}$ matrix is diagonal.
Using these symmetric forms in (\ref{eq:W_analog1}) yields\begin{equation}
\mathbf{P}\mathbf{V^{\prime}}\mathbf{V}^{-1}-\mathbf{P}\mathbf{U}^{\prime}\mathbf{U}^{-1}=\left(\mathbf{U}^{T}\right)^{-1}\mathbf{C}\mathbf{V}^{-1},\label{eq:sym_form3}\end{equation}
in $\left(a,b\right)$. Note that right-hand side is necessarily a
symmetric matrix due to symmetry of the left-hand side.

\subsection{Green's matrix of the second order operator from the Green's matrix
of the corresponding first order differential equation set\label{sub:Green-mtrx-2-from-1}}

Let's rewrite the second order system given in (\ref{eq:scnd_ord_sys})
more explicitly\[
\left.\begin{array}{ccccccccc}
M_{11,x}y_{1}\left(x\right) & + & V_{12}\left(x\right)y_{2}\left(x\right) & + & \dots & + & V_{1n}\left(x\right)y_{n}\left(x\right) & = & h_{1}\left(x\right),\\
V_{12}\left(x\right)y_{1}\left(x\right) & + & M_{22,x}y_{2}\left(x\right) & + & \dots & + & V_{2n}\left(x\right)y_{n}\left(x\right) & = & h_{2}\left(x\right),\\
\vdots & + & \vdots & + & \vdots & + & \vdots & = & \vdots\\
V_{1n}\left(x\right)y_{1}\left(x\right) & + & V_{2n}\left(x\right)y_{2}\left(x\right) & + & \dots & + & M_{nn,x}y_{n}\left(x\right) & = & h_{n}\left(x\right).\end{array}\right.\]
with the boundary conditions\[
y_{i}\left(a\right)=0,\quad y_{i}\left(b\right)=0.\]
Let's rewrite the system of second order differential equations as
a system of first order equations with the definitions\[
z_{i}\left(x\right)\equiv y_{i}\left(x\right),\qquad z_{n+i}\left(x\right)\equiv\frac{d}{dx}y_{i}\left(x\right).\]
Then, the first order system can be put in the following form;\begin{equation}
\mathbf{z}^{\prime}\left(x\right)=\mathbf{A}\left(x\right)\mathbf{z}\left(x\right)+\mathbf{f}\left(x\right)\label{eq:frst_ord_sys}\end{equation}
where\[
\mathbf{A}\equiv\left[\begin{array}{cccccccc}
0 & 0 & \dots & 0 & 1 & 0 & \dots & 0\\
0 & 0 & \dots & 0 & 0 & 1 & \dots & 0\\
\vdots & \vdots & \ddots & \vdots & \vdots & \vdots & \ddots & \vdots\\
0 & 0 & \dots & 0 & 0 & 0 & \dots & 1\\
-\frac{q_{1}}{p_{1}} & -\frac{V_{12}}{p_{1}} & \dots & -\frac{V_{1n}}{p_{1}} & -\frac{p_{1}^{\prime}}{p_{1}} & 0 & \dots & 0\\
-\frac{V_{21}}{p_{2}} & -\frac{q_{2}}{p_{2}} & \dots & -\frac{V_{2n}}{p_{2}} & 0 & -\frac{p_{2}^{\prime}}{p_{2}} & \dots & 0\\
\vdots & \vdots & \ddots & \vdots & \vdots & \vdots & \ddots & \vdots\\
-\frac{V_{n1}}{p_{n}} & -\frac{V_{n2}}{p_{n}} & \dots & -\frac{q_{n}}{p_{n}} & 0 & 0 & \dots & -\frac{p_{n}^{\prime}}{p_{n}}\end{array}\right],\,\mathbf{z}\equiv\left[\begin{array}{c}
z_{1}\\
z_{2}\\
\vdots\\
z_{n}\\
z_{n+1}\\
z_{n+2}\\
\vdots\\
z_{2n}\end{array}\right],\,\mathbf{f}\equiv\left[\begin{array}{c}
0\\
0\\
\vdots\\
0\\
\frac{h_{1}}{p_{1}}\\
\frac{h_{2}}{p_{2}}\\
\vdots\\
\frac{h_{n}}{p_{n}}\end{array}\right].\]
The boundary conditions can be restated in this notation\begin{equation}
\mathbf{B}_{a}\mathbf{z}\left(a\right)+\mathbf{B}_{b}\mathbf{z}\left(b\right)=\mathbf{o},\label{eq:bc_first_ord_sys}\end{equation}
where $\mathbf{B}_{a}$ and $\mathbf{B}_{b}$ are $2n\times2n$ matrices
in the following form \begin{equation}
\mathbf{B}_{a}=\left[\begin{array}{cc}
\mathbf{I} & \mathbf{O}\\
\mathbf{O} & \mathbf{O}\end{array}\right],\quad\mathbf{B}_{b}=\left[\begin{array}{cc}
\mathbf{O} & \mathbf{O}\\
\mathbf{I} & \mathbf{O}\end{array}\right],\label{eq:Ba_Bb}\end{equation}
where $\mathbf{I}$ is the $n\times n$ identity matrix. 

Green's matrix of the first order system can be defined with the formal
solution\[
\mathbf{z}\left(x\right)=\int_{a}^{b}dt\,\mathbf{G}_{1}\left(x,t\right)\mathbf{f}\left(t\right).\]
Then, the formal solution of the second order system is\[
\mathbf{y}\left(x\right)=\mathbf{U}_{n\times2n}\mathbf{z}\left(x\right)=\int_{a}^{b}dt\,\mathbf{U}_{n\times2n}\mathbf{G}_{1}\left(x,t\right)\mathbf{f}\left(t\right)\]
where \[
\mathbf{U}_{n\times2n}\equiv\left[\begin{array}{cc}
\mathbf{I} & \mathbf{O}\end{array}\right].\]
$\mathbf{f}\left(t\right)$ can be rewritten as\[
\mathbf{f}\left(t\right)=\mathbf{L}_{2n\times n}\mathbf{P}^{-1}\left(t\right)\mathbf{h}\left(t\right),\qquad\mathbf{L}_{2n\times n}\equiv\left[\begin{array}{c}
\mathbf{O}\\
\mathbf{I}\end{array}\right].\]
Then,\[
\mathbf{y}\left(x\right)=\int_{a}^{b}dt\,\mathbf{U}_{n\times2n}\mathbf{G}_{1}\left(x,t\right)\mathbf{L}_{2n\times n}\mathbf{P}^{-1}\left(t\right)\mathbf{h}\left(t\right).\]
Comparing this result with (\ref{eq:formal_soln_second}) yields\begin{equation}
\mathbf{G}\left(x,t\right)=\mathbf{U}_{n\times2n}\mathbf{G}_{1}\left(x,t\right)\mathbf{L}_{2n\times n}\mathbf{P}^{-1}\left(t\right).\label{eq:reln_frst_scnd}\end{equation}
Multiplications with $\mathbf{U}_{n\times2n}$ and $\mathbf{L}_{2n\times n}$
choose up-left $n\times n$ block of the Green's matrix of the first
order system.

The relation between the Green's matrix of the first order system
and the Green's matrix of the second order system is established.
Let's continue with reproducing the result of Cole \cite{Cole} for
the Green's matrix of a first order system\emph{.}

\subsubsection{Green's matrix of a first order differential equation set}

Let's have a {}``generic'' first order differential equation set
in the form of;\begin{equation}
\mathbf{z}^{\prime}\left(x\right)=\mathbf{A}\left(x\right)\mathbf{z}\left(x\right)+\mathbf{f}\left(x\right),\label{eq:gen_frst_order_sys}\end{equation}
where $\mathbf{z}$, $\mathbf{f}$ are $l$ dimensional vectors and
$\mathbf{A}$ is an $l\times l$ dimensional matrix. 

First, assume a particular solution of the form\[
\mathbf{z}_{p}\left(x\right)=\mathbf{W}\left(x\right)\mathbf{g}\left(x\right)\]
where $\mathbf{g}$ is an unknown column vector, and the so called
fundamental matrix, $\mathbf{W}$ is the matrix whose columns are
the $l$ linearly independent solutions of the homogeneous differential
equation set, $\mathbf{z}^{\prime}\left(x\right)=\mathbf{A}\left(x\right)\mathbf{z}\left(x\right)$;
i.e.\[
\mathbf{W}^{\prime}\left(x\right)=\mathbf{A}\left(x\right)\mathbf{W}\left(x\right).\]
Then, putting the guess for the particular solution in the non-homogeneous
equation yields the formal solution for $\mathbf{g}\left(x\right)$
as\[
\mathbf{g}\left(x\right)=\int_{a}^{x}dt\,\mathbf{W}^{-1}\left(t\right)\mathbf{f}\left(t\right).\]
Therefore, the general solution for the generic first order differential
equation set is \begin{equation}
\mathbf{z}\left(x\right)=\mathbf{W}\left(x\right)\int_{a}^{x}dt\,\mathbf{W}^{-1}\left(t\right)\mathbf{f}\left(t\right)+\mathbf{W}\left(x\right)\mathbf{c},\label{eq:gen_sol_frst}\end{equation}
where $\mathbf{c}$ is a constant vector.

Any well-posed boundary condition which yields unique solution to
the boundary value problem can be put in a matrix form. Consider the
boundary conditions\begin{equation}
\mathbf{B}_{a}\mathbf{z}\left(a\right)+\mathbf{B}_{b}\mathbf{z}\left(b\right)=\mathbf{o}.\label{eq:mtrx_bc}\end{equation}
Applying these boundary conditions to the general solution fixes $\mathbf{c}$
as\[
\mathbf{c}=-\mathbf{D}^{-1}\mathbf{B}_{b}\mathbf{W}\left(b\right)\int_{a}^{b}dt\,\mathbf{W}^{-1}\left(t\right)\mathbf{f}\left(t\right)\]
where $\mathbf{D}$ is defined by\[
\mathbf{D}\equiv\mathbf{B}_{a}\mathbf{W}\left(a\right)+\mathbf{B}_{b}\mathbf{W}\left(b\right).\]
Using this result in the general solution, one obtained\begin{align}
\mathbf{z}\left(x\right) & =\mathbf{W}\left(x\right)\int_{a}^{x}dt\,\mathbf{W}^{-1}\left(t\right)\mathbf{f}\left(t\right)-\mathbf{W}\left(x\right)\mathbf{D}^{-1}\mathbf{B}_{b}\mathbf{W}\left(b\right)\int_{a}^{b}dt\,\mathbf{W}^{-1}\left(t\right)\mathbf{f}\left(t\right).\label{eq:mtrx_soln_stp1}\end{align}
This result can be put in a final form by rearranging the first term
on the right as;\begin{align*}
\mathbf{W}\left(x\right)\int_{a}^{x}dt\,\mathbf{W}^{-1}\left(t\right)\mathbf{f}\left(t\right) & =\mathbf{W}\left(x\right)\mathbf{D}^{-1}\mathbf{D}\int_{a}^{x}dt\,\mathbf{W}^{-1}\left(t\right)\mathbf{f}\left(t\right),\\
 & =\mathbf{W}\left(x\right)\mathbf{D}^{-1}\left\{ \mathbf{B}_{a}\mathbf{W}\left(a\right)+\mathbf{B}_{b}\mathbf{W}\left(b\right)\right\} \int_{a}^{x}dt\,\mathbf{W}^{-1}\left(t\right)\mathbf{f}\left(t\right),\\
 & =\mathbf{W}\left(x\right)\mathbf{D}^{-1}\mathbf{B}_{a}\mathbf{W}\left(a\right)\int_{a}^{x}dt\,\mathbf{W}^{-1}\left(t\right)\mathbf{f}\left(t\right)\\
 & \phantom{===}+\mathbf{W}\left(x\right)\mathbf{D}^{-1}\mathbf{B}_{b}\mathbf{W}\left(b\right)\int_{a}^{x}dt\,\mathbf{W}^{-1}\left(t\right)\mathbf{f}\left(t\right).\end{align*}
Using this in (\ref{eq:mtrx_soln_stp1}) yields\[
\mathbf{z}\left(x\right)=\mathbf{W}\left(x\right)\mathbf{D}^{-1}\mathbf{B}_{a}\mathbf{W}\left(a\right)\int_{a}^{x}dt\,\mathbf{W}^{-1}\left(t\right)\mathbf{f}\left(t\right)-\mathbf{W}\left(x\right)\mathbf{D}^{-1}\mathbf{B}_{b}\mathbf{W}\left(b\right)\int_{x}^{b}dt\,\mathbf{W}^{-1}\left(t\right)\mathbf{f}\left(t\right),\]
or,\[
\mathbf{z}\left(x\right)=\int_{a}^{b}dt\,\mathbf{G}_{1}\left(x,t\right)\mathbf{f}\left(t\right)\]
where Green's matrix is given as\begin{equation}
\mathbf{G}_{1}\left(x,t\right)\equiv\begin{cases}
-\mathbf{W}\left(x\right)\mathbf{D}^{-1}\mathbf{B}_{b}\mathbf{W}\left(b\right)\mathbf{W}^{-1}\left(t\right), & x<t,\\
\phantom{-}\mathbf{W}\left(x\right)\mathbf{D}^{-1}\mathbf{B}_{a}\mathbf{W}\left(a\right)\mathbf{W}^{-1}\left(t\right), & x>t.\end{cases}\label{eq:Greens_matrix}\end{equation}

\subsubsection{Green's matrix of the second order system}

Using (\ref{eq:reln_frst_scnd}), Green's matrix for the second order
system is\begin{equation}
\mathbf{G}\left(x,t\right)=\begin{cases}
-\mathbf{U}_{n\times2n}\mathbf{W}\left(x\right)\mathbf{D}^{-1}\mathbf{B}_{b}\mathbf{W}\left(b\right)\mathbf{W}^{-1}\left(t\right)\mathbf{L}_{2n\times n}\mathbf{P}^{-1}\left(t\right), & x\le t,\\
\phantom{-}\mathbf{U}_{n\times2n}\mathbf{W}\left(x\right)\mathbf{D}^{-1}\mathbf{B}_{a}\mathbf{W}\left(a\right)\mathbf{W}^{-1}\left(t\right)\mathbf{L}_{2n\times n}\mathbf{P}^{-1}\left(t\right), & x\ge t.\end{cases}\label{eq:Greens_mtrx_second}\end{equation}
Note that $\mathbf{W}$ which is the fundamental matrix of the first
order set is the Wronskian matrix of the second order set. With this
result, we actually achieve our goal which is to construct the Green's
matrix of the second order, self-adjoint matrix differential operator
from the linearly independent solutions of the corresponding differential
equation set. However, we continue to work on this result in order
to find a more compact form and to find forms in which properties
of the Green's matrix are more transparent.

First of all, let's transform Green's matrix to a form in which boundary
conditions is explicit. In order to achieve this goal, let's assume
that the $2n$ linearly independent solutions forming Wronskian matrix,
$\mathbf{W}$ is chosen in such a way that $n$ of them satisfy one
boundary condition, and the other $n$ satisfy the other boundary
condition; i.e.\begin{align*}
\mathbf{M}_{x}\mathbf{u}^{\alpha}\left(x\right) & =\mathbf{o},\quad\mathbf{u}^{\alpha}\left(a\right)=\mathbf{o},\\
\mathbf{M}_{x}\mathbf{v}^{\beta}\left(x\right) & =\mathbf{o},\quad\mathbf{v}^{\beta}\left(b\right)=\mathbf{o},\end{align*}
where $\alpha,\beta=1,\dots,n$. Then, let's choose $\mathbf{W}$
in the form

\[
\mathbf{W}=\left[\begin{array}{cc}
\mathbf{U} & \mathbf{V}\\
\mathbf{U}^{\prime} & \mathbf{V}^{\prime}\end{array}\right],\]
where $\mathbf{U}$ and $\mathbf{V}$ are $n\times n$ matrices whose
columns are $\mathbf{u}^{\alpha}$ and $\mathbf{v}^{\beta}$, respectively.
Using this form of $\mathbf{W}$ greatly simplifies the matrix multiplications
given in (\ref{eq:Greens_mtrx_second}) and yields the results\begin{align*}
\mathbf{D}^{-1}\mathbf{B}_{b}\mathbf{W}\left(b\right) & =\left[\begin{array}{cc}
\mathbf{I} & \mathbf{O}\\
\mathbf{O} & \mathbf{O}\end{array}\right],\qquad\mathbf{D}^{-1}\mathbf{B}_{a}\mathbf{W}\left(a\right)=\left[\begin{array}{cc}
\mathbf{O} & \mathbf{O}\\
\mathbf{O} & \mathbf{I}\end{array}\right].\end{align*}
Block inverse of Wronskian matrix can be given as (See Sect.\ref{sec:Appendix-a:-Inverse});\[
\mathbf{W}^{-1}=\left[\begin{array}{cc}
\mathbf{U}^{-1}+\mathbf{U}^{-1}\mathbf{V}\left(\mathbf{V}^{\prime}-\mathbf{U}^{\prime}\mathbf{U}^{-1}\mathbf{V}\right)^{-1}\mathbf{\mathbf{U}^{\prime}}\mathbf{U}^{-1} & -\mathbf{U}^{-1}\mathbf{V}\left(\mathbf{V}^{\prime}-\mathbf{U}^{\prime}\mathbf{U}^{-1}\mathbf{V}\right)^{-1}\\
-\left(\mathbf{V}^{\prime}-\mathbf{U}^{\prime}\mathbf{U}^{-1}\mathbf{V}\right)^{-1}\mathbf{U}^{\prime}\mathbf{U}^{-1} & \left(\mathbf{V}^{\prime}-\mathbf{U}^{\prime}\mathbf{U}^{-1}\mathbf{V}\right)^{-1}\end{array}\right],\]
where each matrix is a function of $t$. Note that $\mathbf{U}\left(t\right)$
is singular at $x=t=a.$ Thus, $x\le t$ part of the Green's matrix
should be written in such a way that $x=a$ boundary value is given
separately.

Putting these results in (\ref{eq:Greens_mtrx_second}) yields a form
where boundary conditions are explicitly satisfied:\begin{equation}
\mathbf{G}\left(x,t\right)=\begin{cases}
\mathbf{O}, & a=x\le t,\\
\mathbf{U}\left(x\right)\mathbf{U}^{-1}\left(t\right)\mathbf{V}\left(t\right)\left(\mathbf{V}^{\prime}\left(t\right)-\mathbf{U}^{\prime}\left(t\right)\mathbf{U}^{-1}\left(t\right)\mathbf{V}\left(t\right)\right)^{-1}\mathbf{P}^{-1}\left(t\right), & a<x\le t,\\
\mathbf{V}\left(x\right)\left(\mathbf{V}^{\prime}\left(t\right)-\mathbf{U}^{\prime}\left(t\right)\mathbf{U}^{-1}\left(t\right)\mathbf{V}\left(t\right)\right)^{-1}\mathbf{P}^{-1}\left(t\right), & x\ge t.\end{cases}\label{eq:Green_mtrx_3row}\end{equation}
Or, after rearranging, one has a more symmetric form;\begin{equation}
\mathbf{G}\left(x,t\right)=\begin{cases}
\mathbf{O} & a=x\le t,\\
\mathbf{U}\left(x\right)\mathbf{U}^{-1}\left(t\right)\left[\mathbf{P}\left(t\right)\left(\mathbf{V}^{\prime}\left(t\right)\mathbf{V}^{-1}\left(t\right)-\mathbf{U}^{\prime}\left(t\right)\mathbf{U}^{-1}\left(t\right)\right)\right]^{-1}, & a<x\le t,\\
\mathbf{V}\left(x\right)\mathbf{V}^{-1}\left(t\right)\left[\mathbf{P}\left(t\right)\left(\mathbf{V}^{\prime}\left(t\right)\mathbf{V}^{-1}\left(t\right)-\mathbf{U}^{\prime}\left(t\right)\mathbf{U}^{-1}\left(t\right)\right)\right]^{-1}, & b>x\ge t,\\
\mathbf{O} & b=x\ge t.\end{cases}\label{eq:Green_mtrx_4row}\end{equation}

With the help of (\ref{eq:sym_form3}), a final compact form of the
Green's matrix is obtained as \begin{equation}
\mathbf{G}\left(x,t\right)=\begin{cases}
\mathbf{U}\left(x\right)\left(\mathbf{C}^{T}\right)^{-1}\mathbf{V}^{T}\left(t\right), & x\le t,\\
\mathbf{V}\left(x\right)\mathbf{C}^{-1}\mathbf{U}^{T}\left(t\right), & x\ge t.\end{cases}\label{eq:Green_mtrx_fnl}\end{equation}
Or, writing in terms of the elements;\begin{equation}
G_{ij}\left(x,t\right)=\left(\mathbf{C}^{-1}\right)_{\beta\alpha}\begin{cases}
u_{i}^{\alpha}\left(x\right)v_{j}^{\beta}\left(t\right), & x\le t,\\
v_{i}^{\beta}\left(x\right)u_{j}^{\alpha}\left(t\right), & x\ge t.\end{cases}\label{eq:Green_mtrx_elmnt}\end{equation}
where there is summation on the Greek indices.

In all of the above forms of the Green's matrix, some properties are
explicit, while the others not. Now, let's investigate these properties.

\subsubsection{Verifying the properties of the Green's matrix}

Let's show that the Green's matrix that we have constructed satisfies
the properties listed in Section \ref{sec:Properties-of-Green's}:
\begin{itemize}
\item It obviously satisfies the homogeneous boundary conditions in the
forms starting with (\ref{eq:Green_mtrx_3row}).
\item It's columns are formed by the solutions of the homogeneous differential
equation.
\item It is continuous at $x=t$, which is explicit in (\ref{eq:Green_mtrx_3row})
and (\ref{eq:Green_mtrx_4row}).
\item In forms (\ref{eq:Green_mtrx_fnl}) and (\ref{eq:Green_mtrx_elmnt}),
symmetry property is explicit. Let's write $\mathbf{G}^{T}\left(t,x\right)$;
\[
\mathbf{G}^{T}\left(t,x\right)=\begin{cases}
\mathbf{V}\left(x\right)\mathbf{C}^{-1}\mathbf{U}^{T}\left(t\right), & t\le x,\\
\mathbf{U}\left(x\right)\left(\mathbf{C}^{-1}\right)^{T}\mathbf{V}^{T}\left(t\right), & t\ge x,\end{cases}\]
which is simply equal to $\mathbf{G}\left(x,t\right)$.
\item Derivative property of the Green's matrix can be given in the matrix
form as;\begin{equation}
\underset{\epsilon\rightarrow0}{\lim}\left.\frac{d}{dx}\mathbf{G}\left(x,t\right)\right|_{t-\epsilon}^{t+\epsilon}=\mathbf{P}^{-1}\left(t\right).\label{eq:matrx_derv_prop}\end{equation}
Using the Green's matrix form given in (\ref{eq:Green_mtrx_4row});
\begin{multline*}
\mathbf{V}^{\prime}\left(t\right)\mathbf{V}^{-1}\left(t\right)\left[\mathbf{P}\left(t\right)\left(\mathbf{V}^{\prime}\left(t\right)\mathbf{V}^{-1}\left(t\right)-\mathbf{U}^{\prime}\left(t\right)\mathbf{U}^{-1}\left(t\right)\right)\right]^{-1}\\
-\mathbf{U}\left(x\right)\mathbf{U}^{-1}\left(t\right)\left[\mathbf{P}\left(t\right)\left(\mathbf{V}^{\prime}\left(t\right)\mathbf{V}^{-1}\left(t\right)-\mathbf{U}^{\prime}\left(t\right)\mathbf{U}^{-1}\left(t\right)\right)\right]^{-1}=\mathbf{P}^{-1}\left(t\right).\end{multline*}

\end{itemize}

\subsection{Construction with a Guess\label{sub:Construct-guess}}

Now, let's try to construct the Green's matrix in the similar way
as in the case of a single differential operator. A somewhat similar
derivation was given in \cite{Baacke 1997} for a specific case. Since
the columns of the Green's matrix are solutions of the homogeneous
differential equation, Green's matrix should have the following form
in order to satisfy the boundary conditions;\[
\mathbf{G}\left(x,t\right)=\begin{cases}
\mathbf{U}\left(x\right)\mathbf{S}\left(t\right), & x<t,\\
\mathbf{V}\left(x\right)\mathbf{T}\left(t\right), & x>t,\end{cases}\]
where $\mathbf{S}\left(t\right)$ and $\mathbf{T}\left(t\right)$
are unknown matrices. In this form, each column is a linear combination
of the linearly independent solutions. Surely, this form is the most
general guess satisfying the boundary conditions and the homogeneous
differential equation set. After putting the proposal in the derivative
property in (\ref{eq:matrx_derv_prop}), one obtains\[
\mathbf{V}^{\prime}\left(t\right)\mathbf{T}\left(t\right)-\mathbf{U}^{\prime}\left(t\right)\mathbf{S}\left(t\right)=\mathbf{P}^{-1}\left(t\right).\]
Using (\ref{eq:sym_form3}), one has\[
\mathbf{P}^{-1}=\left(\mathbf{V}^{\prime}\mathbf{V}^{-1}-\mathbf{U^{\prime}}\mathbf{U}^{-1}\right)\mathbf{V}\mathbf{C}^{-1}\mathbf{U}^{T}.\]
Putting it in the derivative property (\ref{eq:matrx_derv_prop})
and using the symmetry of $\mathbf{V}\mathbf{C}^{-1}\mathbf{U}^{T}$
yield,\[
\mathbf{V}^{\prime}\mathbf{T}-\mathbf{U}^{\prime}\mathbf{S}=\mathbf{V}^{\prime}\mathbf{C}^{-1}\mathbf{U}^{T}-\mathbf{U^{\prime}}\left(\mathbf{C}^{-1}\right)^{T}\mathbf{V}^{T}.\]
Then, the unknown $\mathbf{S}$ and $\mathbf{T}$ matrices are\begin{align*}
\mathbf{S} & =\left(\mathbf{C}^{-1}\right)^{T}\mathbf{V}^{T},\\
\mathbf{T} & =\mathbf{C}^{-1}\mathbf{U}^{T}.\end{align*}
This result yields the same form given in (\ref{eq:Green_mtrx_fnl}).

\section{Discussion and Conclusion}

In this paper, we constructed the Green's matrix of a second order,
self-adjoint matrix differential operator. This construction is useful
especially in the numerical studies, since obtaining the linearly
independent solutions of the corresponding homogeneous differential
equation set is easy with linearly independent initial conditions.
Just after obtaining the set of linearly independent solutions, one
may directly use (\ref{eq:Greens_mtrx_second}) and obtain the Green's
matrix without considering the boundary behavior of the solutions.
However, once the linearly independent solutions are redefined such
that half of them satisfy one homogeneous boundary condition and the
other half the other boundary, or are obtained directly in this way
with proper boundary conditions, then the compact form of the Green's
matrix in (\ref{eq:Green_mtrx_fnl}) can be used by calculating the
constant matrix in (\ref{eq:W_analog1}). By taking this route, one
may avoid numerical errors coming from correspondingly greater amount
of matrix multiplications and inversions. Also, in cases where solving
the corresponding homogeneous differential equation set analytically
is easy, construction of the Green's matrix by use of (\ref{eq:Green_mtrx_fnl})
may become as easy as other techniques.

Extracting Green's matrix of a higher order differential equation
set from the corresponding first order differential equation set is
a useful technique which is not well known in physics literature.
Although we construct the Green's matrix of a second order, self-adjoint
matrix differential operator by using the Green's matrix of the corresponding
first order differential equation set due to its physical relevance,
Green's matrix for any higher order linear (matrix) operator, either
having self-adjointness property or not, can be extracted from the
Green's matrix of the corresponding first order differential equation
set.

A final comment on the boundary conditions is that a differential
equation set satisfying boundary conditions other than the homogeneous
ones can be handled as in the case of the single differential operator.
Construction of a Green's matrix satisfying boundary conditions other
than homogeneous ones can be handled by the first method given in
this paper.

\section{Appendix A: Inverse of a Block Matrix\label{sec:Appendix-a:-Inverse}}

We reproduce the derivation given by Thornburg \cite{Thornburg}.
Let's have a $2n\times2n$ matrix in the form\[
\left[\begin{array}{cc}
\mathbf{A} & \mathbf{B}\\
\mathbf{C} & \mathbf{D}\end{array}\right]\]
where $\mathbf{A}$, $\mathbf{B}$, $\mathbf{C}$, $\mathbf{D}$ are
$n\times n$ matrices. Inverse of this matrix can be determined by
obtaining the block decomposition of this matrix.

Inverse of the following matrix forms can be found easily: \[
\left[\begin{array}{cc}
\mathbf{A} & \mathbf{O}\\
\mathbf{O} & \mathbf{D}\end{array}\right]^{-1}=\left[\begin{array}{cc}
\mathbf{A}^{-1} & \mathbf{O}\\
\mathbf{O} & \mathbf{D}^{-1}\end{array}\right],\quad\left[\begin{array}{cc}
\mathbf{O} & \mathbf{B}\\
\mathbf{C} & \mathbf{O}\end{array}\right]^{-1}=\left[\begin{array}{cc}
\mathbf{O} & \mathbf{C}^{-1}\\
\mathbf{B}^{-1} & \mathbf{O}\end{array}\right],\]
\[
\left[\begin{array}{cc}
\mathbf{I} & \mathbf{B}\\
\mathbf{O} & \mathbf{I}\end{array}\right]^{-1}=\left[\begin{array}{cc}
\mathbf{I} & -\mathbf{B}\\
\mathbf{O} & \mathbf{I}\end{array}\right],\quad\left[\begin{array}{cc}
\mathbf{I} & \mathbf{O}\\
\mathbf{C} & \mathbf{I}\end{array}\right]^{-1}=\left[\begin{array}{cc}
\mathbf{I} & \mathbf{O}\\
-\mathbf{C} & \mathbf{I}\end{array}\right].\]
If we block decompose a general matrix in such a way that the above
forms appear, then taking inverse can be handled by using the above
relations. In order to obtain the block decomposition, one can use
the following equation\[
\left[\begin{array}{cc}
\mathbf{A} & \mathbf{B}\\
\mathbf{C} & \mathbf{D}\end{array}\right]\left[\begin{array}{c}
\mathbf{E}\\
\mathbf{F}\end{array}\right]=\left[\begin{array}{c}
\mathbf{G}\\
\mathbf{H}\end{array}\right]\Rightarrow\begin{array}{c}
\mathbf{A}\mathbf{E}+\mathbf{B}\mathbf{F}=\mathbf{G},\\
\mathbf{C}\mathbf{E}+\mathbf{D}\mathbf{F}=\mathbf{H}.\end{array}\]
This equation set can be solved for $\mathbf{F}$ by multiplying the
first equation with $-\mathbf{C}\mathbf{A}^{-1}$ and summing with
the second one. These operations are equal to multiplying coefficient
matrix with\[
\left[\begin{array}{cc}
\mathbf{I} & \mathbf{O}\\
-\mathbf{C}\mathbf{A}^{-1} & \mathbf{I}\end{array}\right]\]
from left. Then, coefficient matrix becomes\[
\left[\begin{array}{cc}
\mathbf{A} & \mathbf{B}\\
\mathbf{O} & \mathbf{D}-\mathbf{C}\mathbf{A}^{-1}\mathbf{B}\end{array}\right]\]
where $\mathbf{S}_{A}\equiv\mathbf{D}-\mathbf{C}\mathbf{A}^{-1}\mathbf{B}$
is called Schur complement of $\mathbf{A}$. In order to solve equations
in $\mathbf{E}$, the second equation should be multiplied with $-\mathbf{B}\mathbf{S}_{A}^{-1}$
and summed with the first equation. These operations are equal to
multiplying the modified coefficient matrix with\[
\left[\begin{array}{cc}
\mathbf{I} & -\mathbf{B}\mathbf{S}_{A}^{-1}\\
\mathbf{O} & \mathbf{I}\end{array}\right]\]
from left. Afterwards, the coefficient matrix becomes\[
\left[\begin{array}{cc}
\mathbf{A} & \mathbf{O}\\
\mathbf{O} & \mathbf{S}_{A}\end{array}\right].\]
Thus,\[
\left[\begin{array}{cc}
\mathbf{I} & -\mathbf{B}\mathbf{S}_{A}^{-1}\\
\mathbf{O} & \mathbf{I}\end{array}\right]\left[\begin{array}{cc}
\mathbf{I} & \mathbf{O}\\
-\mathbf{C}\mathbf{A}^{-1} & \mathbf{I}\end{array}\right]\left[\begin{array}{cc}
\mathbf{A} & \mathbf{B}\\
\mathbf{C} & \mathbf{D}\end{array}\right]=\left[\begin{array}{cc}
\mathbf{A} & \mathbf{O}\\
\mathbf{O} & \mathbf{S}_{A}\end{array}\right]\]
which yields the block decomposed form\[
\left[\begin{array}{cc}
\mathbf{A} & \mathbf{B}\\
\mathbf{C} & \mathbf{D}\end{array}\right]=\left[\begin{array}{cc}
\mathbf{I} & \mathbf{O}\\
\mathbf{C}\mathbf{A}^{-1} & \mathbf{I}\end{array}\right]\left[\begin{array}{cc}
\mathbf{I} & \mathbf{B}\mathbf{S}_{A}^{-1}\\
\mathbf{O} & \mathbf{I}\end{array}\right]\left[\begin{array}{cc}
\mathbf{A} & \mathbf{O}\\
\mathbf{O} & \mathbf{S}_{A}\end{array}\right].\]
Then, the inverse of the $2n\times2n$ matrix can be found as\[
\left[\begin{array}{cc}
\mathbf{A} & \mathbf{B}\\
\mathbf{C} & \mathbf{D}\end{array}\right]^{-1}=\left[\begin{array}{cc}
\mathbf{A}^{-1}+\mathbf{A}^{-1}\mathbf{B}\mathbf{S}_{A}^{-1}\mathbf{C}\mathbf{A}^{-1} & -\mathbf{A}^{-1}\mathbf{B}\mathbf{S}_{A}^{-1}\\
-\mathbf{S}_{A}^{-1}\mathbf{C}\mathbf{A}^{-1} & \mathbf{S}_{A}^{-1}\end{array}\right].\]

Using the above result, the inverse of the Wronskian matrix is;\[
\left[\begin{array}{cc}
\mathbf{U} & \mathbf{V}\\
\mathbf{U}^{\prime} & \mathbf{V}^{\prime}\end{array}\right]^{-1}=\left[\begin{array}{cc}
\mathbf{U}^{-1}+\mathbf{U}^{-1}\mathbf{V}\left(\mathbf{V}^{\prime}-\mathbf{U}^{\prime}\mathbf{U}^{-1}\mathbf{V}\right)^{-1}\mathbf{\mathbf{U}^{\prime}}\mathbf{U}^{-1} & -\mathbf{U}^{-1}\mathbf{V}\left(\mathbf{V}^{\prime}-\mathbf{U}^{\prime}\mathbf{U}^{-1}\mathbf{V}\right)^{-1}\\
-\left(\mathbf{V}^{\prime}-\mathbf{U}^{\prime}\mathbf{U}^{-1}\mathbf{V}\right)^{-1}\mathbf{U}^{\prime}\mathbf{U}^{-1} & \left(\mathbf{V}^{\prime}-\mathbf{U}^{\prime}\mathbf{U}^{-1}\mathbf{V}\right)^{-1}\end{array}\right]\]

Note that linear independence of the solutions of the corresponding
homogeneous differential equation set implies invertibility of the
Wronskian matrix and its four elements.

\section{\label{ackno} Acknowledgments}

We thank J.$\!$ Baacke and F.$\!$ Öktem for leading us to useful
references. Preliminary version of this work was presented in {}``$8^{\text{th}}$
Workshop on Quantization, Dualities \& Integrable Systems, Ankara''
on April 23, 2009 by TÇ\c{S}. BT is partially supported by T{Ü}B\.{I}TAK
Kariyer Grant 104T177. TÇ\c{S} is supported by T{Ü}B\.{I}TAK PhD
Scholarship.

\end{document}